\documentclass[preprint,showpacs,preprintnumbers,amsmath,amssymb,endfloats*]{revtex4}
\usepackage{graphicx}

\begin{document}

\thispagestyle{empty}
\title{
Comment on ``On the temperature dependence of the Casimir effect''
}
\author{V.~B.~Bezerra,${}^{1}$
R.~S.~Decca,${}^{2}$ 
E.~Fischbach,${}^{3}$
B.~Geyer,${}^{4}$
  G.~L.~Klimchitskaya,${}^{5,4}$
D.~E.~Krause,${}^{6,3}$
D.~L\'{o}pez,${}^{7}$
V.~M.~Mostepanenko,${}^{8,4}$
and C.~Romero${}^{1}$
}

\affiliation{
${}^{1}$Department of Physics, Federal University of Para\'{\i}ba,
C.P.5008, CEP 58059--970, Jo\~{a}o Pessoa, Pb-Brazil \\
${}^{2}$Department of Physics, Indiana
University-Purdue
University Indianapolis, Indianapolis, Indiana 46202, USA \\
${}^{3}$Department of Physics, Purdue University, West Lafayette, Indiana
47907, USA \\
${}^{4}$Institute for
Theoretical Physics, Leipzig University, Augustusplatz 10/11,
04109, Leipzig, Germany\\
${}^{5}$North-West Technical University, Millionnaya St. 5, St.Petersburg,
191065, Russia\\
${}^{6}$Physics Department, Wabash College, Crawfordsville, Indiana 47933,
USA \\
${}^{7}$Bell Laboratories, Lucent Technologies, Murray Hill, New
Jersey 07974, USA \\
${}^{8}$Noncommercial Partnership ``Scientific Instruments'', 
Tverskaya St. 11, Moscow, 103905, Russia
}

\begin{abstract}
Recently, I.~Brevik, J.~B.~Aarseth, J.~S.~H{\o}ye, 
and K.~A.~Milton [Phys. Rev. E {\bf 71}, 056101 (2005)]
adduced arguments against the traditional approach to the thermal
Casimir force between real metals and in favor of one of the
alternative approaches. The latter assume zero contribution from
the transverse electric mode at zero frequency in qualitative
disagreement with unity as given by the thermal quantum field
theory for ideal metals. Those authors  claim that their approach 
is consistent with experiments as well as with thermodynamics.
We demonstrate that these conclusions are incorrect.
We show specifically that their results are contradicted by four
recent experiments and also violate the third law of thermodynamics
(the Nernst heat theorem).
\end{abstract}

\pacs{42.50.Pq, 03.70.+k, 11.10.Wx, 78.20.Ci}

\maketitle

The paper \cite{1} is devoted to an important problem which has been 
actively discussed in the literature starting in 2000, and  which has 
created much controversy among various research groups. 
The authors of Ref.~\cite{1}
try to justify an alternative approach to the theoretical description
of the thermal Casimir force, which was first suggested in
Ref.~\cite{2} and later supported and further developed in their
own Refs.~\cite{3,4,5}. 
(Henceforth, this approach will be called BSBAHM after the
principal authors Bostr\"{o}m, Sernelius, Brevik, Aarseth,
H{\o}ye, and Milton.)
However, in Ref.~\cite{1} no attention is paid to the serious
shortcomings of this approach, and 
the traditional  approaches in their discussion are misrepresented.

The authors of Ref.~\cite{1}
claim that their calculations of the thermal Casimir 
force between a Cu plate and an Au sphere are:
 a) consistent with current experiments (their wording is 
``not inconsistent'');
 b) consistent with the third law of thermodynamics.
Below we demonstrate that both of these conclusions are incorrect.

a) The first main claim contained in Ref.~\cite{1} is that the
approach by BSBAHM is consistent with current experiments 
(Sec.~III). In Ref.~\cite{11} the opposite conclusion was
drawn, namely that this approach is 
experimentally excluded. 
Note that Ref.~\cite{11} 
contains the results of two experiments performed using
a micromechanical torsional oscillator: a static  
measurement of the Casimir force between a sphere and a plate, and a 
dynamic measurement of the effective Casimir pressure between 
two plane parallel plates. The static measurement was less 
precise than the dynamic measurement over a wide separation range. The 
conclusion that the BSBAHM approach is inconsistent with the experimental 
data in the separation region from $z=260\,$nm to $z=700\,$nm was made in 
Ref.~\cite{11} on the basis of the dynamic measurement. 
This conclusion is illustrated in Fig.~12
of Ref.~\cite{11} where the differences between the theoretical
$P_C^{{th},1}$ (calculated in the framework of BSBAHM approach)
and experimental $P_C^{exp}$ values of the Casimir pressure are
plotted as a function of separation for one set of measurements
containing 235 experimental points. For comparison, in 
Fig.~11 of Ref.~\cite{11} the same differences are
plotted for the theoretical Casimir pressures 
calculated in the framework of the impedance approach of 
Ref.~\cite{12}. 
(Recall that this approach, as well as the approach using the dielectric
function of the plasma model, are called ``traditional'' since they
yield results in qualitative agreement with the case of ideal metals;
the thermal corrections predicted by the alternative approach
of Refs.~\cite{1,2,3,4,5} are many times greater at short
separations.)
From these two figures it becomes apparent that the
BSBAHM approach is excluded by experiment, whereas the impedance
approach is consistent with experimental data.

Surprisingly, in order to 
demonstrate the consistency of their approach with current 
experiments, the authors of Ref.~\cite{1} discuss at length only 
the experiments 
of Refs.~\cite{13,14} and the static measurement of 
Refs.~\cite{11,15}. 
These are at present not the most accurate,
and were not used in the literature to exclude their approach. As to 
the dynamic measurement of Ref.~\cite{11} 
(which contradicts the BSBAHM approach), only a brief mention of this
experiment is made at the end of Sec.~III,
even though this is the most precise current 
measurement. Ref.~\cite{1} claims without proof that there are systematic
theoretical and experimental uncertainties in Ref.~\cite{11} 
connected, in particular,
with a systematic shift of position as discussed earlier
in Refs.~\cite{5,15d}. It is easy to verify, however, that the
reasoning of Refs.~\cite{5,15d} is incapable of avoiding the
conclusion of Ref.~\cite{11} that the BSBAHM approach is excluded
experimentally. According to Refs.~\cite{5,15d}, even a small
experimental error $\Delta z$ in separation distances between two
parallel plates ($\Delta z=1\,$nm in Ref.~\cite{11}) leads to an error
in the theoretical Casimir pressure given by 
$\Delta P_C^{th,1}\approx -P_C^{th,1}(4\Delta z/z)$.
At short separations this error may be rather large.
At separations under consideration in Ref.~\cite{11}
it is, however, much less than the discrepancies between the
BSBAHM theory and experiment. Thus, at the 
shortest separation $z=260\,$nm, $\Delta P_C^{th,1}=3.7\,$mPa, 
which compares with the 
5.5\,mPa mean deviation of the BSBAHM prediction from experiment
as shown in Fig.~12 of Ref.~\cite{11}.
At separations of $z=300,\, 400,\, 500\,$, and 600\,nm
the above error is equal to 1.5, 0.36, 0.12, and 0.05\,mPa,
respectively, which should be compared with much larger mean 
deviations between the BSBAHM approach and experiment at these 
separations (5, 2, 0.8, and 0.4\,mPa, respectively). This
demonstrates that the uncertainties in separations discussed in 
Refs.~\cite{5,15d} do not affect the conclusion of Ref.~\cite{11}
that the BSBAHM approach is excluded experimentally.

Note that even the comparison of the static experiment in Sec.~III 
of Ref.~\cite{1} with the BSBAHM approach is incorrect. Only one point 
at a separation $z=200\,$nm in Fig.~3 of Ref.~\cite{15}
is considered. Instead of using the 
original measured force values, Ref.~\cite{1} uses a maximum
(not a mean) difference 
of --1\,pN between the theoretical (as calculated in Ref.~\cite{15}
at zero temperature) 
and experimental values of the force. The authors of Ref.~\cite{1} 
explain this difference by the existence of a thermal correction
equal to 1\,pN. They
compare this 1\,pN with their predicted thermal correction of 
2.56\,pN at $z=200\,$nm and conclude that the result ``is encouraging". 
In fact, however, they have a deviation of 1.56\,pN  between their 
theory and the static measurement instead of a --1\,pN deviation 
between the traditional theory of 
Ref.~\cite{15} and the same measurement. Ref.~\cite{1}, however, 
does not inform the reader 
that in a later publication \cite{11} the preliminary theoretical 
result of Ref.~\cite{15} was recalculated using a more precise 
roughness correction (this is clearly explained in the left column 
on p.2 of Ref.~\cite{11}). It was demonstrated that the static 
measurement of the Casimir force $F_C^{exp}$ is 
in fact in agreement with the theoretical values $F_C^{th}$
given by traditional theory with a more precise roughness correction
(see Fig.~1). Thus, if one removes this misunderstanding, 
Ref.~\cite{1} must reconcile the zero mean deviation between the 
traditional theory and static experiment (see Fig.~1) 
with an extra 2.56\,pN 
thermal correction predicted by the BSBAHM alternative approach. 
This leads to the evident failure of their approach. 

Moreover, the BSBAHM theoretical approach disagrees significantly
\cite{15a,15b} with the first modern measurement of the Casimir force
between Au surfaces of a plate and a spherical lens by means of 
a torsion pendulum \cite{15c}. In Ref.~\cite{15c} the experimental data
were found to be consistent with the theoretical Casimir force between
ideal metals. A net deviation between the Casimir forces at a
temperature $T=300\,$K and at a separation $z=1\,\mu$m, 
computed for ideal metals
and using the BSBAHM approach, is about 25\% of the Casimir
force between ideal metals
(recall that for ideal metals at $z=1\,\mu$m, $T=300\,$K the
thermal correction is equal to only 1.2\% of the zero-temperature force).
Of this deviation, 19\% is due to the large thermal correction
predicted by BSBAHM. In spite
of the fact that the experimental uncertainty in Ref.~\cite{15c}
at 1$\,\mu$m is less than 10\%, the effect predicted by the
BSBAHM approach was not observed.
No mention of this important experiment is made in Ref.~\cite{1}.

Quite recently the dynamic experiment of Ref.~\cite{11} was repeated
(see Ref.~\cite{19a}) with many important improvements, 
including a significant suppression of the surface
roughness on the interacting surfaces, and a decrease by a factor of 1.7
(down to $\Delta z=0.6\,$nm) of the experimental error in the 
measurement of the absolute separations between the zero roughness
levels. 
An improvement in detection sensitivity, together
with a reduction of the coupling between the micromachined oscillator
and the environment, yielded measurements at smaller separations
between the test bodies (160\,nm instead of 260\,nm).

From the new results
in Ref.~\cite{19a} the conclusion was drawn that the BSBAHM approach to
the thermal Casimir force is excluded experimentally in the separation
region from 170\,nm to 700\,nm at 95\% confidence. In the separation
region from 300\,nm to 500\,nm the BSBAHM approach is excluded
experimentally at even higher confidence of 99\% \cite{19a}.
Here we illustrate these conclusions in Fig.~2a where the differences of
the theoretical (calculated in the BSBAHM approach) and experimental
Casimir pressures are plotted versus separation for 14 sets of
measurements containing 4066 experimental points.
By contrast, in Fig.~2b the same differences are plotted in the case 
that the theoretical Casimir pressures are calculated using the impedance
approach. In both figures the solid lines represent the 95\% confidence
interval for the differences between theoretical and experimental
Casimir pressures as a function of separation. It should be
particularly emphasized that this confidence interval takes into
account all experimental and theoretical errors,
including in full measure the previously discussed
error in the Casimir pressures due to experimental errors
in separation distances as suggested in Refs.~\cite{5,15d}. 
The comparison of Figs.~2a,b clearly demonstrates that the
BSBAHM approach is excluded by the improved dynamic experiment
measuring the Casimir pressure, whereas the impedance approach is
in excellent agreement with experiment. The traditional approach
using the dielectric function of the plasma model is also consistent
with the data (see Ref.~\cite{19a} for
more details).

Thus, the authors' \cite{1}
approach is in contradiction 
not only with a dynamic expe\-riment 
by means of a micromechanical oscillator \cite{11} and the torsion pendulum
experiment \cite{15c}
(which they leave out of their discussion), but also with a static experiment
\cite{11}
and an improved dynamic experiment \cite{19a} which
measure the Casimir force and pressure, respectively.
It should be emphasized that this conclusion cannot be refuted by
introducing an unaccounted systematic error in the measurement of
the surface separation which might be present in the experiments
of Refs.~\cite{11,19a} in addition to the ones discussed above. 
The reason is that the influence of
such an error (if it existed) decreases with an increase in
separation whereas the contribution of the thermal correction, as
predicted in the {BSBAHM} approach, increases with separation
at moderate separations.  
Bearing in mind that the
BSBAHM approach significantly disagrees with experiment in a wide
separation range for several different experiments, 
it is easy to check that
no unaccounted fixed systematic error is capable of bringing
this approach into agreement with data within the whole range
of measurements. As an example, in Fig.~3 we present the pressure
differences ($P_C^{th,1}-P_C^{exp}$) versus separation for one
typical set of measurements where all separations are decreased
by 1\,nm as suggested recently in Ref.~\cite{16a}.
From Fig.~3 it is clearly seen that such hypothetical systematic
error is incapable of bringing the data into 
agreement with the BSBAHM 
approach within a wide separation range from 240\,nm to 700\,nm. 
Note also that the first version of preprint \cite{16a} obtains 
slightly larger
magnitudes of the Casimir pressure than those computed in
Ref.~\cite{19a} in the framework of the BSBAHM approach. (The greatest
difference is at $z=160\,$nm, $T=300\,$K where, according to
Ref.~\cite{16a}, $P_C^{th,1}=-1132\,$mPa against --1125.5\,mPa in
Ref.~\cite{19a}.) The reason for this
difference is that Refs.~\cite{1,16a}
used data for $\varepsilon(i\xi)$ from the work of 
Ref.~\cite{7}. The computations in Ref.~\cite{7}   
contain an error in a
conversion coefficient from eV to rad/s 
(on p.313, left column the value of $1.537\times 10^{15}$ 
is used instead of the correct value
$1.51927\times 10^{15}$ \cite{19a}). 
In addition, the computations of  Ref.~\cite{7} 
utilize \cite{LR} slightly
different values of the Au plasma frequency and relaxation parameter
[$\omega_p=9.03\,$eV, $\nu(T=300\,\mbox{K})=0.0345\,$eV] than
were cited in Refs.~\cite{1,16a}, and actually used in Ref.~\cite{19a}
[$\omega_p=9.00\,$eV, $\nu(T=300\,\mbox{K})=0.0350\,$eV]. 
When both these changes are taken into account in the computations
of Refs.~\cite{1,16a}, the result  $P_C^{th,1}=-1125.5\,$mPa
of  Ref.~\cite{19a} is recovered. (We note that
the abovementioned error does not 
invalidate the computations of the reduction factors in Ref.~\cite{7}, 
which were performed up to only two significant figures.)

b) The second main claim contained in Ref.~\cite{1} is that the
approach by BSBAHM is consistent with thermodynamics. 
The calculations of \cite{1} are based on the Lifshitz formula for the
free energy of a fluctuating field \cite{6}. 
The dielectric permittivites of Au and Cu at nonzero imaginary Matsubara
frequencies are taken from Ref.~\cite{7}. The contribution of the 
zero-frequency term is obtained by the substitution of the Drude 
dielectric function along the imaginary frequency axis,
\begin{equation}
\varepsilon(i\xi)=1+\frac{\omega_p^2}{\xi[\xi+\nu(T)]},
\label{eq1}
\end{equation} 
\noindent
into the Lifshitz formula.
This substitution leads to an absence of the zero-frequency
contribution of the transverse electric mode, and,
in application to metals, results in a serious inconsistency with
thermodymanics. The point is that for perfect lattices with no
defects or impurities but with a nonzero relaxation $\nu(T)$ and
finite conductivity for nonzero $T$, 
the Bloch-Gr\"{u}neisen law leads to  $\nu(0)=0$.
(This important property for perfect lattices remains even when 
the effects of electron-electron collisions are included.)
As proven analytically in Ref.~\cite{8}, for such
perfect lattices the approach by BSBAHM leads to a nonzero
entropy of a fluctuating field at zero temperature given by,
\begin{equation}
S(z,0)=\frac{k_B}{16\pi z^2}
\int_{0}^{\infty}ydy\ln\left[1-
\left(\frac{y-\sqrt{\frac{4\omega_p^2z^2}{c^2}+y^2}}{y+
\sqrt{\frac{4\omega_p^2z^2}{c^2}+y^2}}\right)^2e^{-y}
\right]<0,
\label{entropy}
\end{equation}
\noindent
which depends on the parameter of the system under consideration,
i.e., on the separation distance $z$ ($k_B$ is the Boltzmann constant).

In an attempt to avoid this serious problem, the authors of \cite{1}
use a nonzero value of the Drude relaxation 
parameter at zero temperature arising from the presence of impurities. 
They can then obtain in Secs.~IIA and IV a zero value of entropy at 
zero temperature (a result obtained first by Bostr\"{o}m and Sernelius 
\cite{9,10}). This, however, does not solve the inconsistency of the 
BSBAHM approach with thermodynamics, as is claimed in Ref.~\cite{1}, 
because it is still violated for a perfect lattice having finite 
conductivity and nonzero relaxation at any $T>0$. The unstated 
assumption of the authors is that perfect crystals with no defects 
or impurities do not exist and, therefore, that the laws of
thermodynamics need not apply to
them. This assumption is unphysical. Nernst and Planck
formulated their famous theorem for the case of a perfect lattice 
which is truly an equilibrium system. 
Later this theorem was proven in the framework of quantum statistical
physics for any system with a nondegenerate dynamical state of
lowest energy
(see, for instance, Refs.~\cite{10a,20a}). Consequently, it
is valid for both perfect lattices and lattices with impurities.
The violation of the Nernst 
heat theorem for a perfect lattice would lead to the failure of 
the theory of electron-phonon interactions, and to the eventual 
abandonment of much of condensed matter physics, statistical physics
and thermodynamics. 
For this reason the approach advocated by the authors is, in fact, 
in violation of thermodynamics. 

In addition to the above serious problems, Ref.~ \cite{1}  
contains several misleading statements, including 
the following:

1) The authors discuss the so-called Modified Ideal Metal model 
(MIM), but do not mention that it
violates thermodynamics. Namely, for the MIM 
model the free energy in the case of two parallel plates
[Eq.~(3.2) of Ref.~\cite{1}] can be
identically rearranged to the form 
\begin{equation}
   {\cal F}^{\rm MIM}(z,T)={\cal F}^{\rm IM}(z,T)+
\zeta(3)k_BT/(16\pi z^2),
\label{eq2}
\end{equation}
\noindent
where ${\cal F}^{IM}$ is the Casimir free energy for the usual ideal
metal \cite{16},  and $\zeta(z)$
is the Riemann zeta function. By differentiating both sides
of Eq.~(\ref{eq2}) with respect to $T$ one obtains
\begin{equation}
   S^{\rm MIM}(z,T)=S^{\rm IM}(z,T)-\zeta(3)k_B/(16\pi z^2),
\label{eq3}
\end{equation}
\noindent
where $S^{\rm MIM}$ and $S^{\rm IM}$ are the entropies for the MIM and IM,
respectively. Taking into account that for the usual ideal metal
$S^{\rm IM}\to 0$ when $T\to 0$ \cite{17}, we come to the conclusion
\begin{equation}
   S^{\rm MIM}(a,0)=-\zeta(3)k_B/(16\pi z^2) < 0,
\label{eq4}
\end{equation}
\noindent
i.e., the MIM model violates the Nernst heat theorem. Remarkably the 
result for real metals obtained in Ref.~\cite{1} coincides with 
that for MIM (which violates thermodynamics) 
at large separations and 
does not coincide with the classical limit based on Kirchhoff's
law \cite{22a}.

2) In Sec.~IV of Ref.~\cite{1}
 it is claimed that ``a transverse electric zero mode ... 
should not be present according to Maxwell's equations of 
electromagnetism". This is not correct. Maxwell's equations 
alone do not lead to a contribution of the 
transverse electric mode at zero frequency,
unless they are supplemented by an adequate characterization of the 
material boundaries. More
importantly, the characterization by means of the Drude dielectric
function, used in Ref.~\cite{1}, is inadequate to describe
virtual photons.

3) In Sec.~VI the authors repeat their argument of Ref.~\cite{4} 
that the exact impedances, which depend on transverse 
momentum, lead to a zero contribution of the transverse 
electric mode to the Casimir force, as does the Drude dielectric 
function. 
Ref.~\cite{1} claims that Refs.~\cite{8,12}, where the nonzero
contribution to this mode was obtained,  completely disregard 
the transverse momentum dependence. This, however, is incorrect. In 
Ref.~\cite{8} the dependence of the impedance on a transverse momentum 
is considered in detail. As was demonstrated in Ref.~\cite{8}, this 
dependence disappears in the limit of zero frequency if 
the dispersion equation for the determination of photon
proper frequencies is taken into account 
in the boundary conditions, and, as a consequence,
the reflection properties of virtual photons on a classical boundary
coincide with those of real photons. 
The transverse-momentum dependence in the impedance is, however,
preserved at nonzero frequencies and in the reflection
coefficients at all frequencies. Ref.~\cite{1} fails to account
for this important property, thus resulting in a violation of the
Nernst heat theorem. 
Ref.~\cite{16a} recalls that the Lifshitz
formula contains not real frequencies but the imaginary Matsubara
frequencies for which there is no mass-shell equation. 
However, Ref.~\cite{16a} disregards the fact that
the impedance boundary conditions are prior to the Lifshitz formula.
They are imposed on any real or virtual electromagnetic wave of real
frequency. The next step is the derivation of the dispersion
equation and only after the application of the argument principle
do we arrive at the Lifshitz formula with the imaginary Matsubara
frequencies \cite{12}. If we admit that the
reflection properties of virtual photons on a real boundary coincide with
those of real photons, we must impose the dispersion equation on the
impedance function in the boundary condition formulated in terms of
real frequencies. Therefore, the speculations in Ref.~\cite{16a}
that for imaginary frequencies the angle of incidence becomes meaningless 
are misleading.

4) In the end of Sec.~VI of Ref.~\cite{1}, the claim of Ref.~\cite{5}
is repeated against the extrapolation \cite{8,11,12}
of the impedance function
from the infrared region to zero frequency.
No mention is made of
Ref.~\cite{18} containing the justification for this
extrapolation, and demonstrating that the treatment of the 
zero-frequency mode as in Ref.~\cite{5} results once again in a
violation of the Nernst heat theorem.
Preprint \cite{16a} claims that according to Ref.~\cite{19a}
real data cannot be used in the description. Ref.~\cite{19a},
however, uses exactly the same real data and in the same
frequency range as Ref.~\cite{1} does. In fact, the disagreement
between Refs.~\cite{1} and \cite{19a} is only in the method of
extrapolating real data to zero frequency.

To conclude, the theoretical approach of Ref.~\cite{1} 
is excluded by four already performed experiments,
namely, by the measurements of the Casimir force using a
torsion pendulum \cite{15a,15b,15c}, a micromechanical
oscillator \cite{11}, and by using two determinations
of the Casimir pressure in a micromechanical system \cite{11,19a}.
The main results of Ref.~\cite{1} are also in 
contradiction with fundamental physical principles such as the laws
of thermodynamics. 

\section*{Acknowledgments}
The authors are grateful to A.~Lambrecht and S.~Reynaud for
information on the computation details in Ref.~\cite{7}.
V.B.B. and C.R. were supported by CNPq.
R.S.D. acknowledges financial support from the Petroleum Research
Foundation through ACS-PRF No. 37452-G.
The work of E.F. is supported in part by the US Department of
Energy under Contract No DE-AC02-76ER071428.
The work of G.L.K. and V.M.M. was supported by Deutsche
Forschungsgemeinschaft grant 436RUS113/789/0-1.

%%%%%%%%%%%%%%%%%%%%%%%%%%%%%%%%%%%

\begin{figure*}
\vspace*{-1cm}
\includegraphics{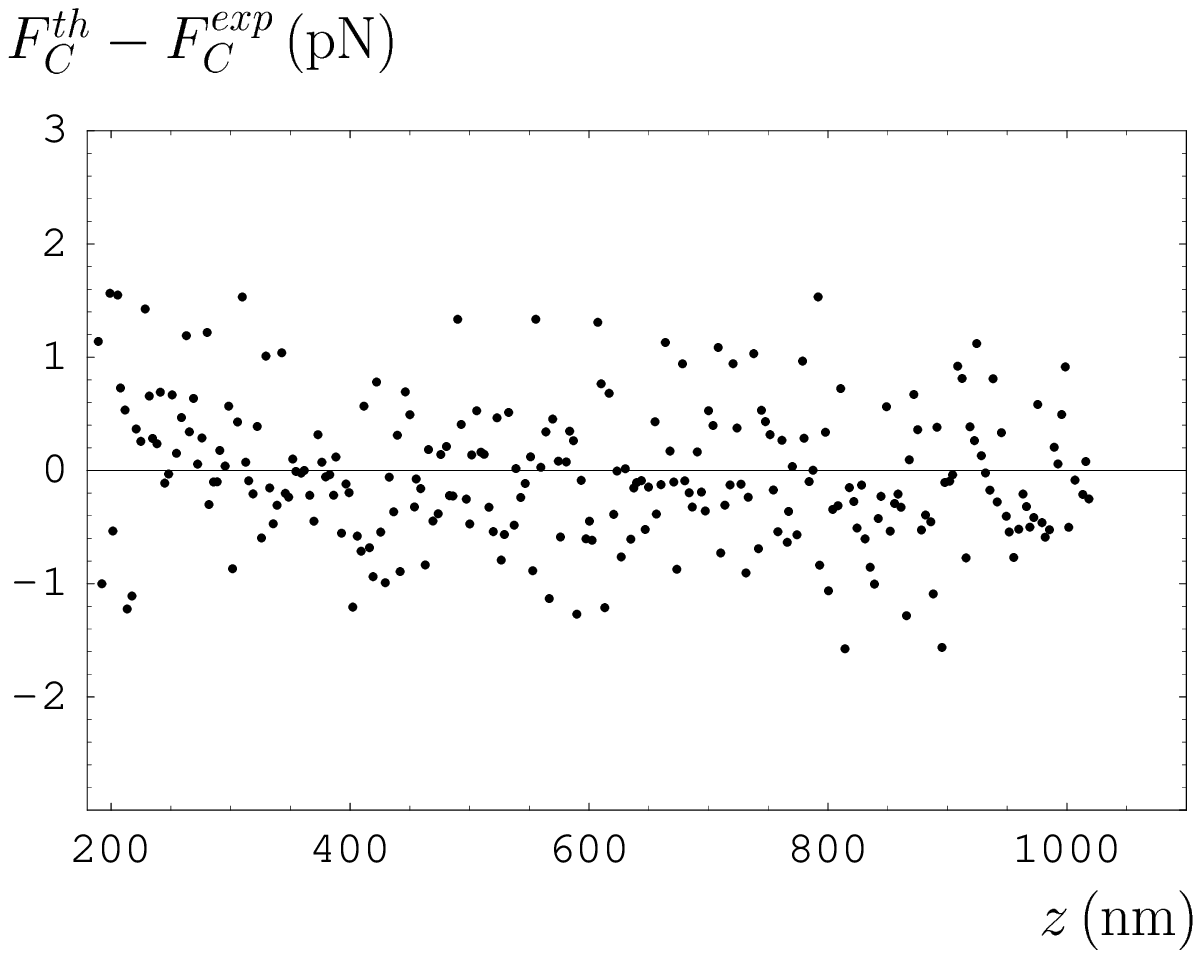}
\vspace*{-8.5cm}
\caption{
Differences of theoretical and experimental Casimir
forces between the sphere and plate versus separation in
the static experiment calculated and measured in Ref.~\cite{11}.
}
\end{figure*}

\begin{figure*}
\vspace*{0.5cm}
\includegraphics{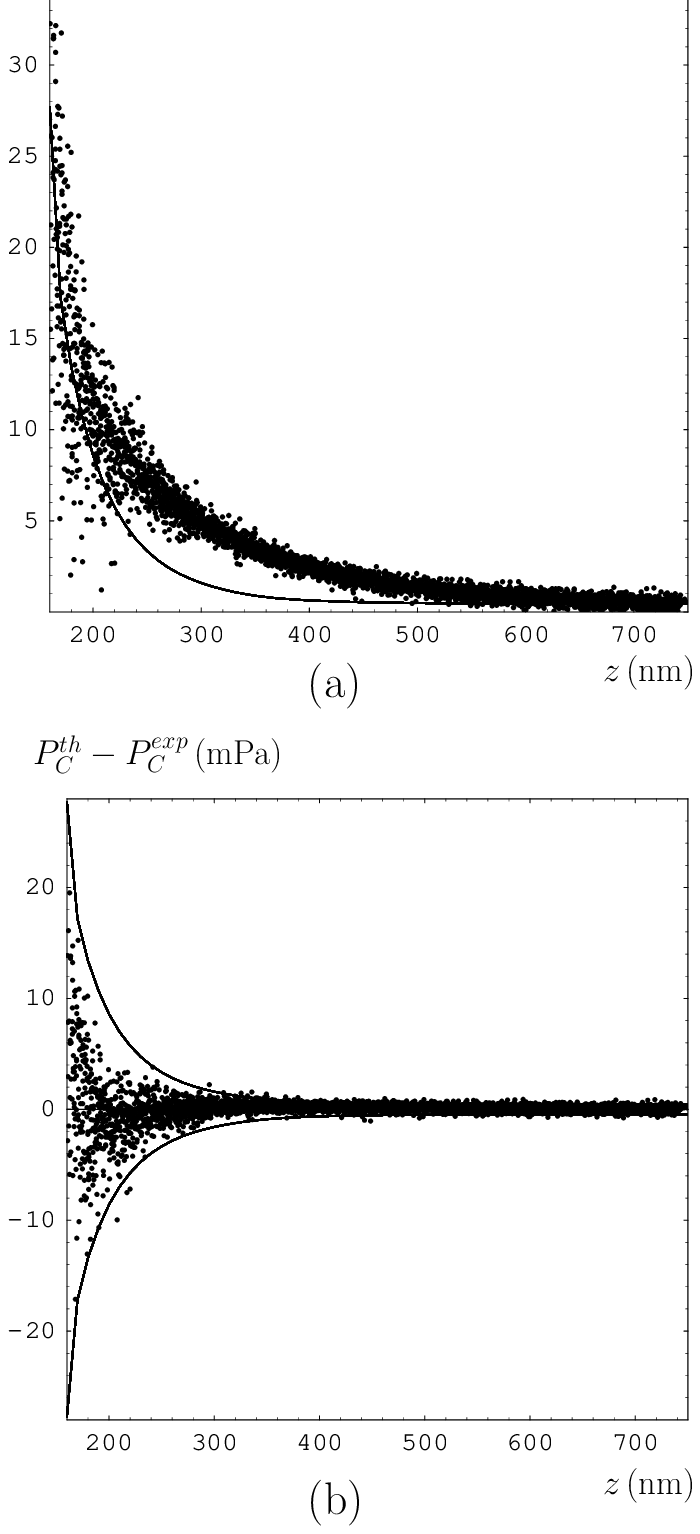}
\vspace*{-12.3cm}
\caption{
Differences of theoretical and experimental 
parallel plate Casimir pressures 
versus separation obtained from the improved dynamic measurement
of Ref.~\cite{19a}.
The theoretical values for $P_C^{th,1}$ are calculated as
in Refs.~\cite{1,2,3,4,5} (a), and for $P_C^{th}$
as in Refs.~\cite{12,19a} (b).
Solid lines represent the 95\% confidence interval.
}
\end{figure*}

\begin{figure*}
\vspace*{-6.8cm}
\includegraphics{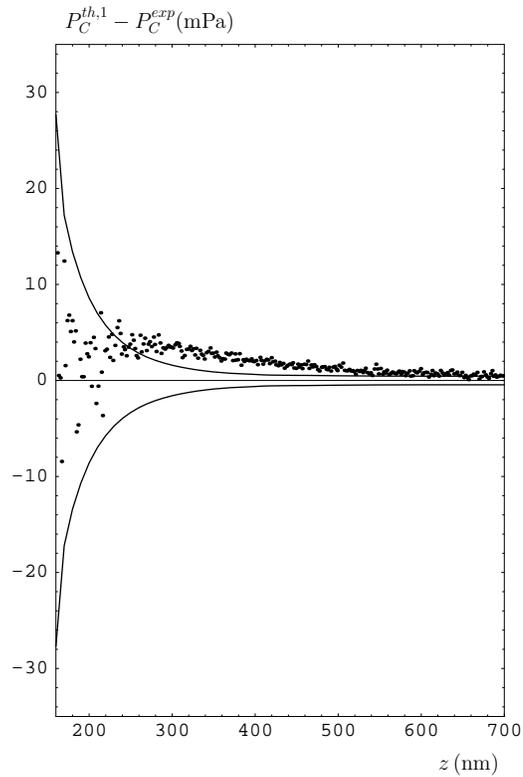}
\vspace*{-10.3cm}
\caption{
Differences of theoretical and experimental 
parallel plate Casimir pressures 
versus separation obtained from the improved dynamic measurement
of Ref.~\cite{19a} after all separation distances 
have been decreased by
1\,nm to take into account the hypothetical systematic error
suggested by Ref.~\cite{16a}.
The theoretical values for $P_C^{th,1}$ are calculated as
in Refs.~\cite{1,2,3,4,5}.
Solid lines represent the 95\% confidence interval.
}
\end{figure*}
\end{document}